\documentclass[10pt,journal,compsoc]{IEEEtran}
\usepackage{graphicx}
\usepackage{amssymb}
\usepackage{multirow}
\usepackage{xcolor}
\usepackage{adjustbox}
\usepackage[para,online,flushleft]{threeparttable}
\usepackage{url}
\usepackage{listings}
\usepackage{caption}

\DeclareCaptionFormat{listing}{#1#2#3}
\ifCLASSOPTIONcompsoc
  \usepackage[nocompress]{cite}
\else
  \usepackage{cite}
\fi
\ifCLASSINFOpdf
\else
\fi
\begin{document}
%
\title{Evaluating Permissioned Blockchain Using Stochastic Modeling and Chaos Engineering}
\thispagestyle{plain}
\pagestyle{plain}

\author{\IEEEauthorblockN{Shiv Sondhi, Sherif Saad and Kevin Shi
}
\IEEEauthorblockA{University of Windsor\\
\{sondhis,shsaad, shi12z\}@uwindsor.ca}
\and
\IEEEauthorblockN{Mohammad Mamun}
\IEEEauthorblockA{National Research Council of Canada\\
mohammad.mamun@nrc-cnrc.gc.ca}
\and
\IEEEauthorblockN{Issa Traore}
\IEEEauthorblockA{University  of Victoria\\
itraore@uvic.ca
}}


%


\maketitle

\begin{abstract}
Blockchain and distributed ledger technologies rely on distributed consensus algorithms. In recent years many consensus algorithms and protocols have been proposed; most of them are for permissioned blockchain networks. However, the performance of these algorithms is not well understood. This paper introduces an approach to evaluating consensus algorithms and blockchain platforms in a hostile network environment with the presence of byzantine and other network failures. The approach starts by using stochastic modeling to model the behaviours of consensus algorithms under different typical and faulty operational scenarios. Next, we implemented a blockchain application using different consensus protocols and tested their performance using chaos engineering techniques. To demonstrate our generic evaluation approach, we analyze the performance of four permissioned blockchain platforms and their consensus protocols. Our results showed that stochastic modeling is an inexpensive and efficient technique for analyzing consensus protocols. But they do not represent the actual performance of the consensus protocols in a production environment. Moreover, an experiment with chaos engineering indicates that if two different blockchain platforms use the same blockchain algorithm or protocol, we should not assume they will have similar performance. Therefore, It is also essential to consider the role of platform architecture and how the protocols are engineered in a given platform.
\end{abstract}


%
\IEEEpeerreviewmaketitle

\section{Introduction}
\label{sec:intro}
Distributed ledger technology (DLT) and blockchain are distributed systems that allow parties who don't fully trust each other to maintain and access a shared database whose state they can always agree on. Although the terms blockchain and DLT are used interchangeably, blockchain is, in fact, a type of DLT where transactions are recorded with an immutable cryptographic signature called a hash. There are many blockchains and DLT platforms available on the market today. They allow users to build and maintain blockchain / DLT systems. They have different architectures and capabilities and offer different tools. In order to choose the optimal platform for a given use case, it is essential to consider these differences. Every DLT has a data-model (eg. linked list, directed acyclic graph), a transaction language, and a consensus protocol or algorithm. However, the choice of each of these components may vary from one implementation to the next, and these choices affect vital system metrics such as performance. For example, the choice of consensus algorithm significantly impacts system performance. Therefore, studying these algorithms is very important.

Deploying a blockchain application directly is the obvious route to measuring its performance, but stochastic modeling techniques are helpful due to their adaptability and time savings. Both stochastic models and blockchain deployments have been used in the existing literature to measure the performance of blockchain consensus algorithms. However, their reliability in hostile environments or in the presence of byzantine and other network failures is not well studied. This research analyzes the usefulness of both methods by conducting experiments with them and comparing the results. We apply chaos engineering techniques to blockchain systems in production by triggering faults on purpose and observing the system's behaviors in a faulty environment. The term chaos engineering was coined in 2014 by Bruce Wong at Netflix, while the practice had already started in 2010 \cite{Basiri16, Tucker18, Torkura19}. Chaos engineering helps to implement fault tolerance strategies that reduce downtime and increase resiliency. The primary motivation for this approach is to overcome uncertainties prevalent in complex computer systems.

Based on the above methodology, we analyze the performance of four permissioned blockchain platforms and their consensus protocols by designing a lightweight blockchain application and implementing it on each platform. The performance metrics evaluated are the load each application can handle, the speed it can serve, and its ability to run as expected. These are important considerations when selecting a blockchain platform. The earliest distributed consensus protocols required each pair of machines on the network to communicate before making a final decision. This resulted in a lower bound of \(O(n^2)\) messages being passed to reach consensus, which did not scale well for large networks. Further research, as in \cite{1_lamport82} and \cite{2_dwork82}, led to the development of faster, more performant protocols. References \cite{3_castro99}, \cite{4_buchman19} and \cite{5_ongaro14} introduce the PBFT, Tendermint, and Raft protocols, respectively, which are evaluated in this paper along with the Clique protocol \cite{20_clique}. The consensus protocol taxonomy provided in \cite{6_sadek20} was used while selecting these protocols and their corresponding blockchain platforms. This is discussed further in Section \ref{sec:consensus}.

Following the aforementioned consensus algorithms, we develop stochastic models and a lightweight application for each model. The chaos engineering principles are then applied to the models and deployments by incorporating different user loads and network faults into the systems. In blockchain applications, faults such as network delay, packet loss, crash failures, and Byzantine failures are injected, while network delay and Byzantine failures are introduced in the models as stochastic elements using probabilistic automata.

{\color{black}{To summarize, we compare stochastic modeling to blockchain deployments by measuring the performance of blockchain consensus protocols in each of them. Models are a step-by-step implementation of the consensus algorithm, whereas deployments are fully functional blockchain applications. In this manner, the performance characteristics of four consensus methods are investigated. The performance of each method is also examined in light of chaos engineering. Our experiments enable us to fairly weigh the pros and cons of building stochastic models over deploying blockchain applications. In addition, they allow us to compare consensus algorithms based on their performance characteristics. The work proposed in this article is an extension of our previous work on evaluating the performance of blockchain consensus protocols \cite{sondhi2021empirical,sondhi2021chaos }

The rest of this paper is organized as follows. Section \ref{sec:related} discusses the existing literature in this domain, Section \ref{sec:consensus} examines our choice of protocols and performance metrics, Section \ref{sec:techniques} describes our methodology, Section \ref{sec:performance} describes our performance evaluation experiments, and Section \ref{sec:results} presents our experiment results. Finally, Section \ref{sec:conc} suggests possible future work and concludes the paper.

\section{Related Work}
\label{sec:related}
Several works in the existing literature measure the performance characteristics of blockchain systems. Some use stochastic modeling tools to model consensus protocols and run simulations, while others deploy a blockchain application using blockchain platforms. Blockchain platforms treat consensus as one component of the application. Other components may include a REST API, the application's business logic, and a front-end, which are not required in the models. For this reason, building, a blockchain application may be a time-consuming process and differs from platform to platform. On the other hand, building a model is quicker and more repeatable. However, blockchain applications are better indicators of performance and security characteristics as they represent the final product. The existing literature relevant to this research is discussed below.

\begin{itemize}
    \item In \cite{11_ampel19}, Ampel et al. measured the performance characteristics of a Hyperledger Sawtooth application using a performance benchmarking tool called Hyperledger Caliper \cite{21_caliper}. The authors used metrics like throughput, latency, success rate, and node resource utilization to evaluate the performance of the Raft consensus protocol used in Sawtooth. {\color{black}{Throughput is the number of transactions committed to the blockchain per unit time, latency is the amount of time it takes for a transaction to be committed to the blockchain since when it was sent, and the success rate is the ratio of successfully committed blocks to the total number of blocks created (including invalid blocks).}} These metrics were plotted against batch size (transactions per block) and the input workload. The throughput increased linearly, and latency increased exponentially with batch size. Latency also increased exponentially with an increasing workload, while memory and CPU usage picked up as well. 

    \item \textcolor{black}{Moschou et. al. conducted experiments to evaluate the performance of two different Hyperledger Sawtooth transaction processors\cite{moschou20}. Their setup consisted of the SETH transaction processor paired with a Node JS RESTful API and the Golang transaction processor paired with the Golang REST API. They used Proof-of-Elapsed Time (PoET) and a random-leader protocol called "DEV protocol" for network consensus and selected execution time as their performance metric. The Golang processor outperformed SETH in all experiments. The results also indicated that the DEV protocol and an increase in the number of validators independently resulted in higher execution times.}
    
    \item \textcolor{black}{In \cite{kuzlu19}, Kuzlu et. al. used Hyperledger Caliper to measure the throughput and latency of a Hyperledger Fabric application while varying its input workload i.e. incoming transactions per second. Their results reveal that while the total number of transactions sent to the application does not affect its performance, higher transaction rates (more than 200 transactions per second) and simultaneously arriving transactions cause app performance to degrade. The authors use AWS EC2 instances with 16 vCPUs and 32 GB of RAM.}

    \item In \cite{8_hao18}, Hao et al. compared Ethereum's PoW to Hyperledger Fabric's PBFT in terms of performance. Average throughput and latency were once again used as comparison metrics. Results indicate that PBFT is better than PoW in terms of both metrics. {\color{black}{For smaller input workloads (around 100 transactions per second), PBFT was only slightly better than PoW, but as the workload increased, PBFT's performance grew far better than that of PoW.}} This is an indication of the poor scalability of the lottery-based PoW consensus. 
    
    \item In \cite{10_asgaonkar18}, Asgaonkar et al. used a Poisson process to build a PoW consensus protocol model. In a Poisson process, two events must occur independently of eachother, must not occur simultaneously, and their average rate of occurrence must be constant. Here, the events in question are the syncing of local blockchain copies on the network. The metrics used were throughput and the number of orphan blocks, where orphan blocks are defined as blocks proposed by a peer which do not appear on any other peer's local blockchain. The authors plotted these metrics against a varying rate of growth between the number of nodes, and the Poisson process parameter, \(\lambda\). 
    
    \item Ilja et al. compared the BFT-based consensus mechanism in Bitfury’s Exonum framework, to Bitcoin's proof-of-work (PoW) in \cite{7_ilja19}. They used the Modest toolset to model the protocol, using packet error rate (PER) as the stochastic component, and recording the minimum and maximum block commit times. Commit times were plotted against a varying PER, to study their correlation. The authors also verified the two-thirds majority voting principle of \cite{1_lamport82}. The results indicate that increasing the PER leads to an exponential increase in time to commit to the BFT protocol, but leads to forks and higher latency in the Bitcoin blockchain. Forks occur in a blockchain when two or more nodes create a valid block almost simultaneously. This leads to inconsistent copies of the chain at different nodes, which may cause nodes to provide contradicting information about the blockchain state. The authors noted that comparing the two protocols was not easy due to their differing finalties. Finalty is the amount of time it takes for a block to be permanently added to the chain.
    
    \item Piriou et al. analyzed the performance of the BizCoin cryptocurrency protocol (a vote-based protocol) in terms of consistency, and its ability to discard double spending attacks \cite{9_piriou18}. The authors proposed three consistency metrics - consensus probability (probability that all processes agree on the same blockchain state), consistency rate (a mean portion of the network that agrees on the most common blockchain state), and worst process delay (the length difference between the main blockchain and its greatest common prefix\footnote{A chain, \(c1\), is the prefix of chain \(c2\), if the last block of \(c1\) is an ancestor of the last block of \(c2\). For the greatest common prefix, \(c2\) is the main-chain, and \(c1\) is its longest prefix that is present in every node's local blockchain.}). The authors built a model using a tool called pyCATSHOO \cite{19_pycatshoo21} and tracked the metrics over time by running simulations. The results showed that consensus probability gradually degraded by 50\%, whereas the worst process delay degraded exponentially by 50\%. The consistency rate degraded only slightly. The authors also used Markov chains to track the probability of the blockchain being in a safe state (no double spending).
    
    \item In \cite{ahmad21} Ahmad et al. compared five different protocols based on transaction throughput and latency. They measured these metrics whilst varying the number of network nodes. The protocols used were PoW, PoS, Proof-of-Elapsed Time (PoET), Clique (Proof-of-Authority) and PBFT. They found that Clique and PoS experienced the minimum latency, followed by PoET, PoW, and PBFT. In terms of throughput they found that with up to 50 network nodes, Clique achieved the best throughput followed by PoET and PoS, but when nodes were increased beyond 50, Clique's throughput degraded. PBFT had a low throughput. 
    
    \item \textcolor{black}{The performance of 4 permissioned blockchain platforms - Hyperledger Fabric, Corda, Quorum, and Ethereum - are measured by Monrat et. al. in \cite{monrat20}. Each platform uses a different consensus protocol - Raft, custom protocol, QuorumChain (PBFT) and PoA respectively. The performance metrics used were throughput and latency. Hyperledger Fabric registered the best performance, followed by Corda, Quorum, and Ethereum. Unlike in \cite{kuzlu19}, the performance of Hyperledger Fabric was not adversely affected by the input transaction rate since Monrat et. al. built the Fabric app with a world state database.}
    
    \item In \cite{angelis18}, Angelis et al. studied Aura, Clique (variants of the Proof-of-Authority class of consensus algorithms), and classical PBFT, using the CAP (Consistency, Availability, Partition tolerance) theorem principles. The CAP theorem states that a distributed system  cannot achieve consistency and availability when the network is partitioned in a way that messages may be arbitrarily lost. In a blockchain network, consistency refers to all nodes having the same blockchain copy, and availability refers to the network's ability to accept new transactions. Through a qualitative analysis, the authors showed that Aura and Clique tend to prefer availability while PBFT prefers consistency.
    
    \item The objective of Duan et al. in \cite{12_duan18} was to provide a reproducible methodology for formal verification of blockchain systems. They achieved this through hierarchical and modular SDL (Specification and Description Language) models, using a private crowdfunding blockchain application. The focus was on the security and safety of blockchain systems. The authors outlined the steps to build blockchain models and emulate malfunctioning nodes. They also provided formal descriptions wherever possible. 
    
    \item The work done by Gopalan et al. in \cite{13_gopalan20} revolved around stability and scalability analysis of blockchain systems using modeling techniques. Stability was defined as the ability of a blockchain to be consistent across peers, for short bursts of time, infinitely many times. Scalability was defined as the property of a blockchain being stable for a given burst-length, as the peers increase monotonically. Through a highly technical analysis, the authors showed how "one-endedness" is desirable for blockchain protocols as it relates directly to the network having no forks and consequently a successful consensus protocol. The experiments were conducted using simulation and real data from the Bitcoin blockchain. 
    
    \item In \cite{14_papadis18}, Papadis et al. used modeling techniques to analyze the block generation statistics of a blockchain system. They compared the results using a blockchain application and a simulated model. They also analyzed the impact of stochastic components on the probability of attacks on the network. The Ethereum testbed was used for building the application, and the difficulty parameter, hashing power of nodes, and network delays were varied as stochastic elements. The authors found that the probability of a successful attack increased with increasing delay and decreased with a higher number of transaction confirmations. 
    
    \item \textcolor{black}{Zhang et. al. propose ChaosETH - a chaos engineering tool for resilience assessment of Ethereum clients\cite{zhang21}. The experiments conducted are used to verify that ChaosETH is able to identify metrics that remain stable during chaos injection, that ChaosETH can give valuable insights into the resilience of a client and that it is able to identify common error models between two Ethereum clients. The experiments were conducted on Go Ethereum and Open Ethereum by invoking system call errors and measuring 15 application-level metrics in all.} 
\end{itemize}

\begin{table*}[htbp]
\caption{Comparing this work to the existing literature}\label{tab:lit}
\footnotesize
\centering
\begin{adjustbox}{width=1\textwidth}
\begin{threeparttable}

\begin{tabular}{|@{\vrule width0ptheight9pt\enspace}l|c|c|c|c|c|c|c|c|}\hline

\hfil\bf Paper&\bf Modelling&\bf Application& \bf Protocol Families &\bf Performance Metrics&\bf Load / Chaos Testing\\\hline

\hfil \cite{11_ampel19} &No &Yes &Paxos\cite{lamport98}\cite{lamport01} &TP\tnote{1}, L\tnote{2}, SR\tnote{3}, RU\tnote{4} &Load\\\hline
\hfil \cite{kuzlu19} &No &Yes &Raft\cite{5_ongaro14} &TP, L &Load\\\hline
\hfil \cite{8_hao18} &No &Yes &PoW\cite{satoshi08}, BFT\cite{1_lamport82} &TP, L &Load\\\hline
\hfil \cite{10_asgaonkar18} &Yes &No &PoW &TP, OR\tnote{5} &None\\\hline
\hfil \cite{7_ilja19} &Yes &No &PoW, BFT &L &Load\\\hline
\hfil \cite{9_piriou18} &Yes &No &N/A (vote-based) &Consistency &Load\\\hline
\hfil \cite{ahmad21} &No &Yes &PoW, BFT, PoS\cite{pos20}, PoA\cite{poa18}, PoET\cite{poet18}\tnote{6} &TP, L &Load\\\hline
\hfil \cite{monrat20} &No &Yes &Raft, N/A (custom), PoA, PBFT\cite{3_castro99} &TP, L &Load\\\hline
\hfil \cite{angelis18} &No &No &BFT, PoA &CAP Theorem &None\\\hline
\hfil \cite{13_gopalan20} &Yes &Yes &PoW &Scalability, Stability &Load\\\hline
\hfil \cite{14_papadis18} &Yes &Yes &Ethereum (i.e. PoW or PoA) &Block generation &Load\\\hline
\hfil This work &Yes &Yes &BFT, PoS, PoA, Paxos &TP, L, SR, Consistency &Both\\\hline

\end{tabular}
\begin{tablenotes}
\item[1] Throughput; \item[2] Latency;
\item[3] Success Rate; \item[4] Node Resource Utilization;
\item[5] Orphan Block Rate; \item[6] Proof-of-Elapsed Time
\end{tablenotes}
\end{threeparttable}

\end{adjustbox}
\end{table*}

Of the papers discussed above, \cite{11_ampel19}, \cite{8_hao18}, \cite{10_asgaonkar18}, \cite{7_ilja19}, \cite{kuzlu19}, \cite{9_piriou18}, \cite{ahmad21}, \cite{13_gopalan20}, \cite{14_papadis18} and \cite{monrat20} are all empirical studies. \cite{11_ampel19}, \cite{kuzlu19}, \cite{8_hao18}, \cite{ahmad21} and \cite{monrat20} did not use modelling techniques. They measured the performance of one or more protocols using similar metrics which include throughput and latency. Of the others, all works used modeling tools, but only \cite{13_gopalan20} and \cite{14_papadis18} verified their modeling results against blockchain deployment results. Most of the works studied either one or two protocols and used metrics like throughput, latency, rate of creation of orphan blocks, consistency metrics, and block generation statistics. In \cite{angelis18}, Angelis et al. conducted a qualitative analysis of three permissioned consensus protocols using the CAP theorem for analysis. They used neither stochastic models nor blockchain applications.

From the existing literature, it is evident that modeling and open-source blockchain platforms are two prominent ways of studying the characteristics of blockchain systems. Moreover, \cite{13_gopalan20} and \cite{14_papadis18} highlighted the importance of verifying stochastic modeling results to blockchain application results. It is also safe to say that throughput and latency are very popular performance indicators for blockchain applications. However, these two metrics are not good enough alone because they contain no information about the consistency of local chains or about the number of invalid/rejected blocks. Similarly, although the CAP theorem offers valuable insight into the characteristics of a protocol, the findings from a CAP theorem analysis must be backed up by implementing the supporting scenarios with the help of metrics like throughput and length of the blockchain as described in \cite{angelis18}.

In our work, we use throughput, latency, success rate, and the standard deviation of local chain lengths as performance metrics. The success rate is defined as the ratio of accepted blocks to the total number of blocks created (which includes rejected blocks). In addition, using chaos engineering principles we test for load tolerance and fault tolerance of the applications. These metrics are discussed further in Section \ref{sec:consensus}. Finally, our experiments are conducted on stochastic models as well as applications deployed on blockchain platforms. A full comparison between this work and the existing literature is summarized in Table \ref{tab:lit}.


\section{Consensus Protocols and Performance}
\label{sec:consensus}
There are close to a hundred consensus protocols used in blockchain and distributed ledger systems today \cite{families18}. {\color{black}{As per the findings in \cite{15_benchmarking17}, at the time of writing, 15 consensus protocols were used most commonly across several industries. Participants of this study include institutions like IBM, R3, Depository Trust and Clearing Corporation (DTCC), BigChainDB, and banks like BBVA, UBS, and more.}} However, there is no single best protocol - the choice depends on network structure, topology, desired confirmation times, security and other factors. This research focuses on permissioned consensus protocols suitable for industries like healthcare and finance that deal with sensitive and private user information. Most enterprises today prefer permissioned (or private) blockchains. Here, an enterprise refers to any company irrespective of size, that follows a centralized governance model (like a board of directors). They constitute a large majority of all corporations today, while the opposing side is mostly made up of decentralized autonomous organizations (DAOs). Therefore, focusing on consensus protocols used in private settings seems more relevant. However, some protocols in this study (like Tendermint) can be used in a public setting as well.

While selecting consensus protocols for our experiments, the taxonomy from \cite{6_sadek20} was used to cover as many different types of protocols as possible. The structural and performance properties of consensus protocols are most relevant here and are discussed below.

\subsection{Selected Protocols}
The structural properties of consensus algorithms can be divided further into the following subcategories: 
\begin{enumerate}
    \item \textbf{Node type} - depending on the platform, a consensus algorithm may deal with many types of nodes like full nodes (which store the entire blockchain locally), validator nodes, endorsers (which only validate transactions), and light clients (which verify new blocks without storing the entire blockchain locally).
    \item \textbf{Structure type} – Consensus protocols can use single or multiple committees to reach consensus i.e. a single group of validators generates each next block (as in PBFT, Tendermint, and Clique), or multiple committees work independently. Both types can be static or dynamically changing. Furthermore, a single committee may be open or closed to new members, and can have implicit or explicit formation rules. Multiple committee mechanisms must have an overall topology (i.e. flat or hierarchical). {\color{black}{Raft normally follows a single committee structure, but when the network is partitioned, this splits into multiple flat committees. If any partition contains more than two-thirds of the participating nodes, it becomes the main committee and the others must follow its decisions (hierarchical topology).}} 
    \item \textbf{Underlying mechanism} – This refers to the core method of reaching consensus and can roughly be classified as either a \emph{lottery-based} (proof-of-work), \emph{vote-based} (BFT-based protocols) or \emph{coin-age-based} mechanism.
\end{enumerate}

The consensus protocols selected for this research - PBFT, Tendermint, Clique, and Raft - belong to the byzantine fault-tolerant (BFT), proof-of-stake (PoS), proof-of-authority (PoA) and Paxos-based protocol families respectively. 

BFT-based protocols are always byzantine fault tolerant. Usually, they follow multiple rounds of voting to achieve consensus (like PBFT) but this is not necessary. Many BFT-based protocols simply suggest improvements over PBFT, like reducing the number of voting rounds, etc. The second family - PoS-based protocols - use a proof-of-stake model in the consensus mechanism. This is commonly used for leader-elections. For instance, the block proposer in each round can be decided based on the validators' stakes in the system. Interestingly, Tendermint is a DPoS-BFT protocol - it uses a PoS model for leader election, and voting rounds to commit blocks. DPoS stands for Delegated PoS, a variant of PoS where any network participant can delegate their tokens to a validator as a vote of confidence. PoA protocols are a popular class of non-incentivized protocols that store proof of each validator's identity to monitor and limit malicious activity. While PoA protocols are also byzantine fault-tolerant, they can reach better performance than BFT-based protocols due to lighter message exchanges. PoA protocols are best suited to scenarios where the validator set can be trusted, as is the case with Ethereum's Rinkeby, G\"{o}rli and Kovan testnets. Finally, Paxos-based protocols provide improvements over the Paxos protocol proposed in 1989. Raft is a popular Paxos-based protocol which (like Paxos itself) is not byzantine fault-tolerant but crash fault-tolerant. Table \ref{tab:protocols} summarizes the properties of the four selected consensus protocols.

\begin{table*}[tbp]
\caption{Comparing the selected consensus protocols}\label{tab:protocols}
\centering
\begin{adjustbox}{width=1\textwidth}
\begin{tabular}{|@{\vrule width0ptheight9pt\enspace}l|c|c|c|c|c|c|}\hline

\hfil\bf Protocol &\bf Family &\bf Platform & \bf Fault Tolerance &\bf Structure &\bf Underlying Mechanism\\\hline

\hfil PBFT &BFT-based &Hyperledger Sawtooth &BFT &Single Committee &Vote-based\\\hline
\hfil Tendermint &PoS-based (DPoS-BFT) &Tendermint Core, Cosmos SDK &BFT &Single Committee &Vote-based\\\hline
\hfil Clique &PoA-based &Ethereum's Rinkeby testnet &BFT &Single Committee &Leader-follower\\\hline
\hfil Raft &Paxos-based &Hyperledger Fabric &CFT &Single / Multiple Committee &Vote-based\\\hline

\end{tabular}
\end{adjustbox}
\end{table*}

\subsection{Selected Performance Metrics}
Performance metrics are a way to quantify a system's performance. The performance properties of consensus protocols defined in \cite{6_sadek20} include throughput, latency, fault tolerance, scalability, and energy consumption. {\color{black}{For our experiments we used the literature referenced in Table \ref{tab:lit} as well as \cite{6_sadek20} and selected a set of primary metrics which directly measure aspects of the system like transaction throughput, average latency and success rate; and secondary metrics which measure changes after the chaos was introduced into the system.}} In the stochastic models, we computed an additional primary metric i.e. the standard deviation of local chain lengths. These metrics are defined below. 
\medskip

\noindent\textbf{Primary Metrics}
\begin{enumerate}
    \item \textbf{Write Throughput} - The number of transactions added to the blockchain per second. 
    
    \[TP = \frac{(total\ transactions\ added\ to\ chain)}{(total\ runtime)}\]

    \item \textbf{Average Write Latency} - The amount of time it takes for a transaction to appear on the blockchain, from when it was made. We are concerned with the average overall transactions.  
    
    \[L = \frac{\sum_{tx=1}^{TX_{tot}} (T_{txCommitted} - T_{txCreated})}{TX_{tot}}\]
    
    Where $TX_{tot}$ is the total number of transactions, $T_{txCommitted}$ is the timestamp when a given transaction is committed and $T_{txCreated}$ is the timestamp when a given transaction is created i.e. made by the user.

    \item \textbf{Success Rate} - The ratio of the number of blocks successfully added to the blockchain to the total number of blocks created (including invalid blocks). 
    
    \[SR = \frac{(total\ successfully\ added\ blocks)}{(total\ blocks\ created)}\]
    
    \item \textbf{Std. Deviation of Local Chain Lengths} - The standard deviation of the lengths of each node's local blockchain copy. This metric was found to be more useful than the success rate while studying the secondary metrics in the stochastic models. 
    
    \[\sigma = \sqrt{\frac{1}{N} \sum_{i=1}^{N} (x_{i} - \mu)^2}\]
    
    Where $N$ is the total number of nodes in the blockchain network, $x_{i}$ is the length of the blockchain at node $i$ and $\mu$ is the mean blockchain length for all nodes in the network.
\end{enumerate}

\noindent\textbf{Secondary Metrics}
\begin{enumerate}
    \item \textbf{Load Tolerance} - Measured by observing changes in performance under a varying input workload. 

    \item \textbf{Fault Tolerance} - Measured by observing changes in performance when faults appear in the network (delay, crash faults, byzantine faults, etc.).
\end{enumerate}

\section{Experiment Test Bed Construction}
\label{sec:techniques}
Any system built around a blockchain is called a blockchain system. Blockchain systems can be classified into five abstract layers – the data-model, network, consensus, execution, and application layers \cite{8_hao18} and \cite{16_dinh17}. The data-model layer defines the data structures and data types of the data stored on the blockchain. The network layer deals with all network protocols while the consensus layer deals with finding consensus on the network and creating new blocks. Together, these are called the core blockchain layers. The execution layer includes details of the runtime environment used to execute smart contracts\footnote{For example, Ethereum's runtime environment is the Ethereum Virtual Machine (EVM).}. The application layer is the topmost layer and represents decentralized applications (Dapps) that use smart contracts and the blockchain to accomplish some business logic. 

Evidently, consensus protocols are at the heart of the consensus layer. They are a well-defined instruction set, which ensures that all nodes on the network agree on the blockchain state (data-model layer). Therefore, the consensus, network, and data-model layers are tightly knit. This means that changes in the network or data model, like network delays, faulty nodes, corrupted messages, and block size can affect the process of consensus. However, modern protocols are built to overcome, minimize or work with the effects of such variations. This section outlines the creation of our two test beds - the stochastic models and the blockchain test bed. The models work only in the core blockchain layers while the blockchain deployments consist of all 5 layers. The benefits of stochastic models are also discussed in this section, and finally, our experiment parameters are presented.

\subsection{Stochastic Models}
Stochastic modeling is the process of modeling under probabilistic uncertainty. In stochastic systems, the relationship between input and output variables is not deterministic - there is a degree of randomness in output determination. Stochastic models account for this non-determinism in state transitions and can simulate seemingly random occurrences in communication systems (like network delays and node failures). Most of this is done using probability distributions and probabilistic automata. Markov chains work well for performance evaluation tasks. 

Testing systems early in the development lifecycle is a good way to judge system performance and allows developers to change architectural decisions (like the choice of consensus algorithm) while they still can \cite{17_giovanni04}. However, to early-test a blockchain application, the network and application must be deployed as well. On the other hand, a model of the consensus protocol can be used to simulate the entire system. Modeling tools are highly configurable and can save time, effort, and money. This makes them very useful. For instance, most consensus algorithms require the ability to add and extract transactions from a transaction pool, or count votes. Several functions like these can be reused across protocols. Models also allow a wide range of evaluation metrics to be defined as is evident in the existing literature where metrics like throughput, latency, scalability, stability, block generation statistics, and multiple consistency metrics were computed.





In this research, we use continuous-time Markov chains, where the models' state space is discrete and the time parameter is continuous i.e. events do not take place after fixed time intervals. A python library called pyCATSHOO \cite{19_pycatshoo21} is used to build the models and run simulations. Although the pyCATSHOO models follow the respective protocol pseudocodes, they adopt a component view of the network. Three network components - the leader, peers, and clients - are defined and can communicate using message channels and references to external variables. {\color{black}{Code Snippet \ref{fig:code-mod-mc} shows the message channel definitions in code.}} Since message passing between component instances is not supported, a separate counter component is used to count peer votes. The components act as state machines, where events trigger state transitions. {\color{black}{Fig.~\ref{fig:st-tend} shows the state transition diagram for the peer component in Tendermint.}} The state space is \{\emph{Start, Waiting, Propose, Prevote, Precommit}\} and a two-thirds majority vote amongst the validators triggers the transition between the last three states. Additionally, sensitive functions like resetCounters(), addToChain(), etc. are executed when a state is entered or exited, or when a reference variable is updated.  {\color{black}{Figs. \ref{fig:code-mod-aut1} and \ref{fig:code-mod-fn} provide the code that define the peer state machine and the addToChain() sensitive function.}}. 

\begin{figure*}[tbp]
    \centering
    \includegraphics[width=\textwidth]{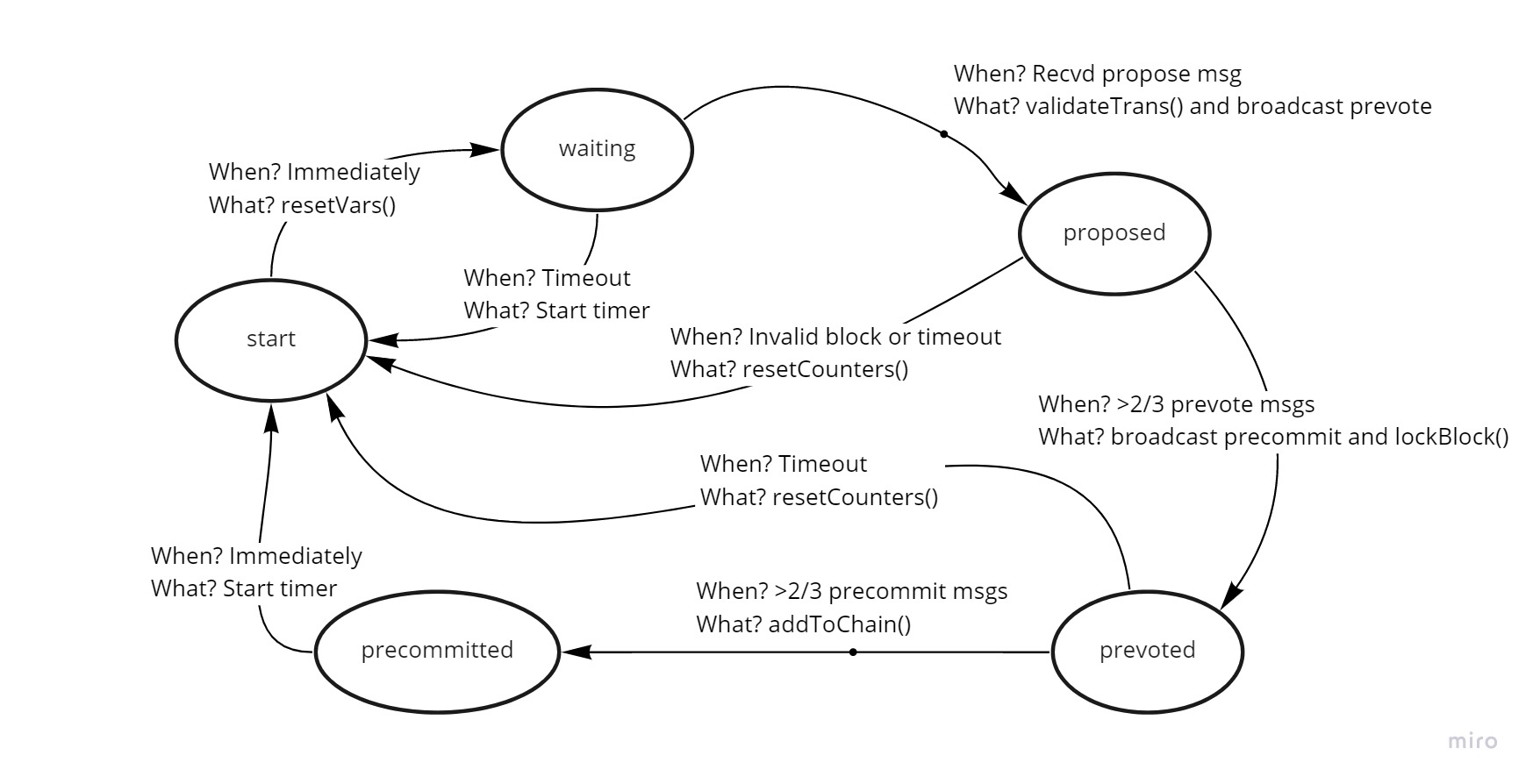}
    \caption{State Transition Diagram of Tendermint's Peer Component}
    \label{fig:st-tend}
\end{figure*}

Stochastic models are clearly not the same as blockchain applications. For example, since the peers are instances of a class and not separate machines, they do not have their own unique resources. Also, message passing is almost instantaneous, which is not representative of real communication scenarios. However, there are two crucial differences. Firstly, as mentioned, votes are counted at a counter rather than at each peer. Secondly, our models do not use a blockchain data structure. This means that when consensus is reached, the new block's number and ID are saved, but the block itself is not saved. Depending on the use case this may not be the best approach, but metrics like throughput, latency, and success rate can easily be measured without having to save individual blocks.

There are two main stochastic elements introduced in the models - time delay and byzantine nodes. Time delay is used to simulate realistic network communication, and byzantine nodes help measure fault tolerance. For delay, each state transition that requires communication amongst peers in the real-world occurs probabilistically using the exponential distribution. For byzantine nodes, each model has a user-defined byzantine rate, which decides what portion of the network peers are byzantine. A byzantine leader will send block proposals with contradicting information to different peers, and a byzantine peer will cast contradictory votes for any block that it receives. {\color{black}{Code Snippet \ref{fig:code-mod-aut2} shows how an automaton defines when and how peers switch between malicious and benign behaviors. Peers always turn malicious in accordance with the byzantine rate and a predefined probability distribution.}} Table \ref{tab:mod-par} lists the important simulation parameters used in the models, along with their values. 

\begin{table}[!htbp]
\caption{Modelling simulation parameters}\label{tab:mod-par}
\centering
\begin{tabular}{|@{\vrule width0ptheight9pt\enspace}l|c|c|}\hline
\textbf{Parameter}&\textbf{Value}\\\hline
    Number of validators&6\\\hline 
    Transactions per block&70\\\hline
    Input transaction rate&1000, 5000, 10000, 15000\\\hline
    Maximum simulation timesteps&500000\\\hline
    Exponential distribution rate parameter, \(\lambda\)&2\\\hline
    Byzantine rate&0, 2\\\hline
\end{tabular}
\end{table}

To summarise, the fundamental difference between our stochastic models and blockchain deployments is that the former model is just the consensus protocol, while the latter is a larger system with consensus as one component. Despite this, modeling can help realize time and energy savings and can help make critical architectural and performance decisions early in the development lifecycle.

\subsection{Blockchain Test Bed}
Each shortlisted consensus algorithm is available on a different platform. Two of these platforms belong to the Hyperledger suite of blockchain technologies, established under the Linux Foundation. For our experiments, we built an application on each platform, using the following business logic: 
\begin{itemize}
    \item User A sends funds worth \emph{x} units to User B. 
    \item User A's account balance is decreased by \emph{x} units. 
    \item User B's account balance is increased by \emph{x} units.  
\end{itemize}

This is a simple asset transfer application. However, depending on the desired use case, the platforms allow for much more functionality, including user registration and a fully functional web application. We selected a simple application in order to obtain results that were representative of the underlying protocols' performance - additional features would result in performance overhead. Below is a description of the blockchain platforms.


\subsubsection{\textbf{Hyperledger Sawtooth}}
Hyperledger Sawtooth \cite{22_sawtooth} uses a modular framework that separates the system's business logic from application-level procedures, making it easier for developers to work with. It supports dynamic consensus i.e. the ability to switch between consensus protocols in-between voting rounds, and pluggable consensus i.e. the ability to choose from a list of protocols. Sawtooth supports Go, Java, JavaScript, and Python SDKs. 

The Sawtooth application was built using version 1.2.6 with the PBFT consensus model. Each node had four docker containers - a REST API endpoint, a consensus engine (PBFT), a validator, and a transaction processor. {\color{black}{Code Snippet \ref{fig:code-saw} provides the docker configuration for the Sawtooth validator.}} The default transaction processor called the intkey transaction processor allows for the creation of an account with an initial balance, modification and listing of an account's balance, and listing of all account balances. For each node, the REST API was exposed in a docker file and was used for communication over the network.

\subsubsection{\textbf{Cosmos-SDK}}
Comos-SDK \cite{23_cosmos} is an open-source framework that supports PoS and PoA blockchain applications. It runs through Tendermint Core - which uses the Tendermint consensus protocol. Tendermint Core deals with the core blockchain layers and comes with an application-blockchain interface (ABCI) which lets it communicate with higher-level tools like Cosmos SDK and other applications. Cosmos provides tools to build a blockchain application and interact with the blockchain itself. The transaction flow in a typical Cosmos application is as follows:
\begin{itemize}
    \item The client sends a message to the app using the CLI or a gRPC endpoint. 
    \item Based on the message type (make a transaction, query blockchain, etc) a message object is created.
    \item This triggers an event that is handled by the handler, a component that consists of functions written to handle specific events. The defined functions typically end with a call to the keeper. 
    \item The keeper is the only component of a Cosmos application that communicates with the blockchain. This is done via the ABCI. Based on the handler's instructions, it can read or write to the blockchain. 
\end{itemize}



The Cosmos application was built using the Launchpad version (v0.39), which has since been updated with breaking changes to Stargate (v0.42). Cosmos provides a scaffolding tool called Starport, which can be used to build a template application, rebuild the app or run it from its last state. The template application comes with 8 pre-built and 1 custom module. The custom module can be modified to fit any business logic. {\color{black}{Each module has its own handler and keeper - the relevant portions of code are provided in Figs. \ref{fig:code-cos-handle} and \ref{fig:code-cos-keep}}}. Accounts and validators are created at runtime with their balances and stakes provided in a configuration file. Cosmos exposes three ports for the application - one each for the Tendermint consensus engine, the REST API, and the application front-end written in Vue.

\subsubsection{\textbf{Go Ethereum}}
Go Ethereum (or geth) \cite{24_geth}, is an Ethereum client written in the Go programming language. Like other implementations of Ethereum, it resides on every node of the network and can run on the Ethereum mainnet as well as the testnets. As a result, geth offers the Ethash protocol (Ethereum's PoW) and the Clique PoA protocol. It works through a JSON-RPC API, and web3 libraries which allow developers to run, maintain, debug and monitor their nodes. Version 1.10.3 of Geth was used in this work. 

The following steps were followed to build the Geth application: 
\begin{itemize}
    \item Create validator accounts (address, password, and key).
    \item Create the genesis block with Clique consensus, block creators, and account balances. 
    \item Compile each node's address into a static node list, which is shared amongst the validators. 
    \item Start all the nodes. 
\end{itemize}

\subsubsection{\textbf{Hyperledger Fabric}}
Hyperledger Fabric \cite{25_fabric} is a permissioned DLT platform, with a modular and highly configurable architecture. The ledger is shared by organizations, each having its own peers and validators (called orderers). Fabric supports Javascript, Go, and Python for writing chain-code, and supports the Raft and Kafka consensus protocols. The transaction flow in a typical Fabric app is as follows:
\begin{itemize}
    \item The client sends a transaction to every organization, that validates it, and sends back an endorsement if valid. 
    \item The client sends the transaction and endorsements to an orderer organization that runs the consensus protocol.
    \item Transactions endorsed by a majority are accepted. 
    \item Once ordered the transactions are sent to the organizations and committed by their peers. 
\end{itemize}

The application was built using Fabric 2.x and the following steps were followed to build it: 
\begin{itemize}
    \item Create the certificate authorities and generate certificates for each organization using Docker and a Fabric binary. 
    \item Register orderers and peers with the organizations and create crypto-material for them.
    \item Generate the genesis block and other channel artifacts. 
    \item Create the peers and orderers, along with their volumes and environments, using Docker. 
    \item Create the channel and join peers to it.
    \item Write the chaincode, install dependencies, package the chaincode, install it at the endorsing peers, and commit it if approved by a majority of the organizations. 
    \item Build the application using Node.js and the Fabric API. 
\end{itemize}


{\color{black}{Figs. \ref{fig:code-fab-ca} and \ref{fig:code-fab-ord} give the docker configurations for the Certificate Authorities and orderers respectively. Code Snippet \ref{fig:code-fab-chan} shows the process of creating and joining a channel, while Code Snippet \ref{fig:code-fab-cc} shows the process of packaging and installing the chaincode. Code Snippet \ref{fig:code-fab-cccfn} shows the relevant chaincode function to add a transaction to the ledger and Code Snippet \ref{fig:code-fab-app} shows the relevant endpoint of the Node.js application which creates a new transaction. \cite{pavan21} was referred while building the Hyperledger Fabric application.}}

\medskip 

\noindent Table \ref{tab:par-app} lists the important parameters used in the application tests, along with their values: 
\begin{table}[!htbp]
\caption{Application test parameters}\label{tab:par-app}
\centering
\begin{tabular}{|c|c|}\hline
\textbf{Parameter}&\textbf{Value}\\\hline
    Number of validators&6\\\hline
    Block size&10 tx/block \emph{or} default in MB\\\hline
    Baseline user load&250, 50\\\hline 
    Load test user loads&250, 500, 1000, 1500\\\hline
    Locust workers&3\\\hline
    Users per second per worker&1, 2\\\hline
\end{tabular}
\end{table}

\section{Chaos Engineering For Blockchain}
\label{sec:performance}
Once the apps are built, their performance is compared using the metrics discussed in Section \ref{sec:consensus}. Further, each protocol is compared on the basis of load testing and chaos engineering, which is used to evaluate the fault tolerance of the applications. This section discusses the tools and methodology used for performance measurement in the stochastic models as well as the blockchain deployments.

In the stochastic models, the primary metrics are measured using simple counters and timers. The results are aggregated and recorded at the end of each simulation. For the applications, a load testing tool called Locust \cite{26_locust}, is used to generate a constant, manageable load on the application, and the metrics are tracked over an entire test run. Locust interacts with the applications using HTTP requests and records the time for a response, the type of response (success or failure), and the total number of successful responses per second. {\color{black}{Code Snippet \ref{fig:code-fab-perf} presents the relevant code excerpt from the Fabric app's locust-file.}}

In the models, the secondary metrics are calculated by varying the input transaction workload (transactions per second) for load tests, and adding stochastic elements as discussed in Section \ref{sec:techniques} for chaos testing. In the applications, Locust is used for load tests, and the load is varied till the application crashes or performance degrades noticeably. For chaos testing, Pumba \cite{27_pumba} is used to generate network delay, loss, and message corruption for relevant network addresses. Pumba is used exclusively with docker containers and uses the traffic control (tc) tool within the Linux iproute2 package under the hood. For applications that do not use docker, like the Geth application where each validator is created on a separate virtual machine, the traffic control (tc) tool is used directly. These methodologies apply for all except the Cosmos application.

Cosmos does not allow HTTP requests to change the blockchain state, instead, HTTP requests can only be used to test the result of a transaction. Therefore, the Cosmos application's performance is measured by using the tmux command line utility in Linux, to open multiple terminal sessions and send requests to the Cosmos app using the command line. This is similar to opening multiple connections (users) to the application. The metrics are then computed using timers and counters and aggregated at the end of each test. {\color{black}{Code Snippet \ref{fig:code-cos-perf1} shows how multiple terminal sessions are opened and Code Snippet \ref{fig:code-cos-perf2} shows how the performance testing file calculates latency and throughput.}} Additionally, since the Cosmos SDK exists at the application layer of the blockchain stack, the Tendermint validators cannot be accessed and faults cannot be injected onto the network\footnote{Some of these functionalities are more readily available in later versions of Cosmos SDK. For example, the validator set can be configured using a docker-compose file in v0.42.}. Therefore, network faults could not be injected and chaos testing was not performed for the Cosmos application.

\section{Discussion Of Results}
\label{sec:results}
\subsection{Stochastic Models}
In the models' throughput, latency, and standard deviation of local chain lengths ($\sigma$), were computed against a varying workload. The metrics were computed four times for each protocol model - with byzantine validator nodes, simulated delay, both delay, and byzantine nodes, and with no stochastic elements (i.e. the baseline). In Figs.~\ref{fig:mod-tp}-\ref{fig:mod-sr}, the performance of the experiments with simulated delays and byzantine nodes are generally flanked by the baseline and the case where a delay and byzantine activity occur together.

\begin{figure*}
  \centering
  \includegraphics[scale=0.7]{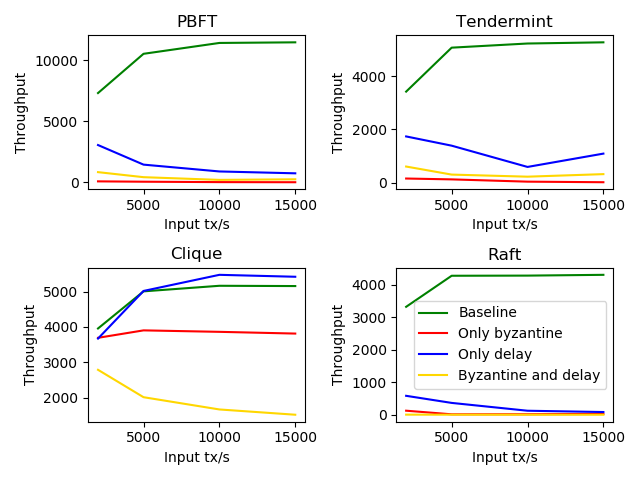}
  \caption{Stochastic models: write throughput}
  \label{fig:mod-tp}
\end{figure*}

\begin{figure*}
  \centering
  \includegraphics[scale=0.8]{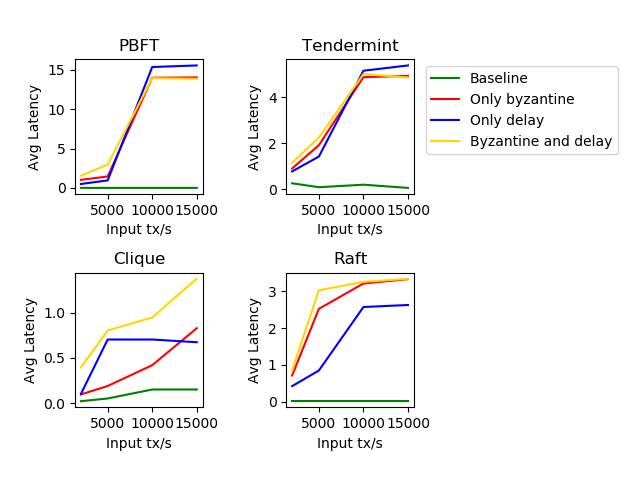}
  \caption{Stochastic models: average latency}
  \label{fig:mod-lat}
\end{figure*}

\begin{figure*}
  \centering
  \includegraphics[scale=0.8]{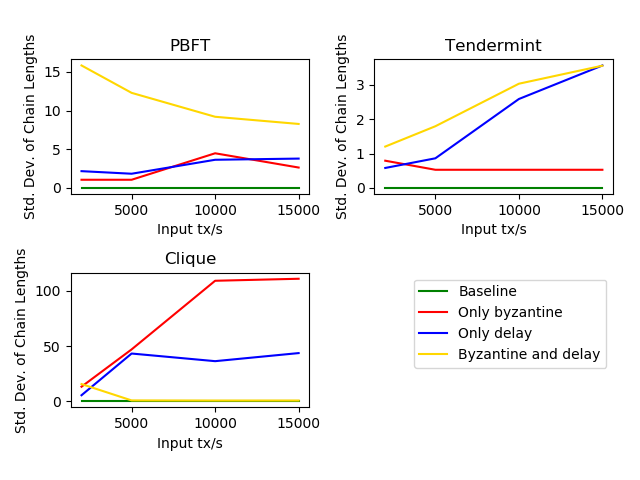}
  \caption{Stochastic models: $\sigma$}
  \label{fig:mod-cl}
\end{figure*}

\begin{figure}
  \centering
  \includegraphics[width=\hsize]{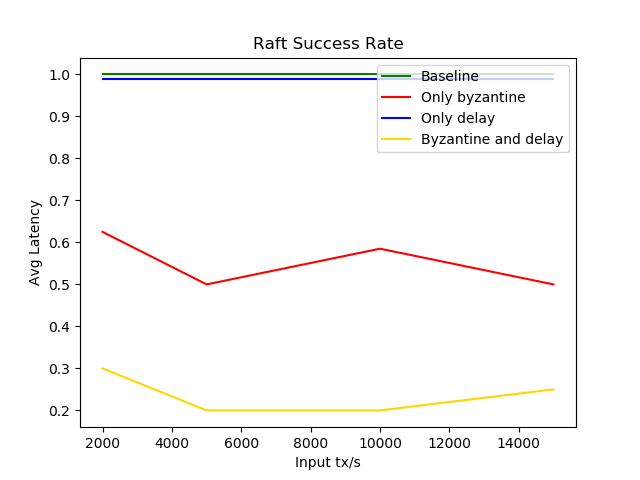}
  \caption{Stochastic models (Raft): success rate}
  \label{fig:mod-sr}
\end{figure}

Fig.~\ref{fig:mod-tp} shows the change in throughput for each protocol as the input transaction workload is varied. For PBFT, Tendermint, and Raft, the baseline throughput is well separated from throughput measurements in the presence of stochastic elements. For the Clique protocol model, adding network delays and byzantine nodes did not affect throughput as much as it did in the other models. This is because of Clique's leader-follower architecture. In the other protocols, when half the network is byzantine, consensus cannot be reached due to contradicting votes being sent across the network. Due to this, system throughput degrades. However, since the peers in Clique do not communicate before adding a block to their local chains, each peer simply accepts or declines the block proposed by the leader based on validity. If the leader itself is byzantine, this may lead to inconsistent local chains, but since a portion of the network receives a valid proposal, the overall throughput of the system does not degrade. Overall, PBFT showed the best baseline performance in terms of throughput while the other three models' results were compared to each other. The addition of byzantine nodes generally affected throughput more than network delays did. 

Fig.~\ref{fig:mod-lat} shows the change in average latency for each protocol as the input transaction workload is varied. The results for the PBFT and Tendermint models were almost identical, although Tendermint had lower latency. As with throughput, latency results in the presence of stochastic elements were similar and well separated from the baseline. The similarity in PBFT and Tendermint latency results is understandable since both protocols use the same BFT-based voting rounds (though Tendermint uses one more round of voting than PBFT). The latency results for Raft also followed a similar pattern to the ones for PBFT and Tendermint, however, there was some separation amongst results in the presence of stochastic elements. For the Clique protocol model, although the results were well separated, in terms of magnitude there was not much difference in results no matter what stochastic elements were added. Interestingly, the average latency for Clique in the presence of byzantine nodes did not flatten out as it did with the other models. 

When simulations were run with byzantine nodes and delay, the success rate for PBFT, Tendermint, and Clique models was different for each local blockchain copy. This is because the nodes receive either contradicting or delayed messages, which results in different nodes reaching different conclusions at the end of each round (no consensus). In other words, some nodes might add a block to their local blockchain while others might not. For this reason, the standard deviation of local chain lengths is used to quantify the inconsistency in lengths amongst the local chains. From Fig.~\ref{fig:mod-cl} it can be seen that for PBFT and Tendermint, adding byzantine failures with network delay caused the local blockchains to diverge the most. For Clique, adding both faults together did not affect the local chains as much. Overall, the addition of stochastic elements affected the crash fault tolerant Clique model more than it did the byzantine fault tolerant models. The spread of chain lengths in the Tendermint model was the smallest, while it was slightly larger for PBFT and considerably larger for Clique. 

Since delay and byzantine nodes did not affect the consistency of local chains for the Raft model, it was left out from Fig.~\ref{fig:mod-cl}. This means that the local chains in the Raft protocol did not diverge during the simulation. However, the presence of byzantine nodes did have an effect on Raft's overall success rate as shown in Fig.~\ref{fig:mod-sr}. Since all local chains were consistent, the success rate at each node was identical and is called the overall success rate. In Fig.~\ref{fig:mod-sr}, the baseline success rate for Raft was equal to its success rate when network delays were simulated, they are separated in the plot for visibility.

\subsection{Blockchain Applications}
In the applications, throughput, latency, and success rate were calculated at a constant input load. We call these the baseline results, which are presented in Table \ref{tab:baseline}. The throughput and latency were also measured while varying the load and while adding faults to the blockchain network. These are called the load and chaos tests, respectively. The load test results are plotted in Figs.~\ref{fig:app-lt-pbft}-\ref{fig:app-lt-raft} and the chaos test results are plotted in Figs.~\ref{fig:app-ft-pbft}-\ref{fig:app-ft-raft}. Table~\ref{tab:ft} presents the chaos testing results by providing the average value for each metric (throughput and latency) while each network fault is being injected into the network.

\begin{table*}[tbp]
\caption{Blockchain applications: baseline performance results}\label{tab:baseline}
\centering
\begin{adjustbox}{width=1\textwidth}
\begin{tabular}{|@{\vrule width0ptheight9pt\enspace}l|c|c|c|c|c|c|c|}\hline
\hfil\bf Protocol&\bf Write Throughput (tx/s)&\bf Avg. Latency (ms)& \bf Success Rate&\bf User Count (Load)\\\hline
\hfil PBFT&50&1100&0.88&250\\\hline
\hfil Tendermint&93.1&2039&1.0&250\\\hline
\hfil Clique&27.3&49&1.0&250\\\hline
\hfil Raft&5.8&1850&0.98&50\\\hline
\end{tabular}
\end{adjustbox}
\end{table*}

The load was generated for the blockchain applications in terms of the number of users interacting with the app, as opposed to the number of input transactions per second for the models. In Table~\ref{tab:baseline}, a manageable load of 250 users was selected for each protocol in order to get as stable results as possible. However, Raft (Hyperledger Fabric) could not deal with a load of 250 users. \textcolor{black}{Kuzlu et. al. ran into a similar problem in \cite{kuzlu19}}. This problem arises due to the way endorsement works in Hyperledger Fabric rather than due to the protocol itself. When a peer validates a transaction in order to give its endorsement, it processes the transaction and obtains the resultant ledger state, called the read set. After the transaction is accepted and ordered while being committed, it is processed once again and the resultant ledger state is called the write set. If the read and write sets do not match the transaction is canceled. This is not ideal for applications expecting large workloads because the state changes several times between endorsement and committing of a block. \textcolor{black}{Since Kuzlu et. al. used a considerably more powerful machine than us, they experienced performance loss at higher transaction rates than we did. Zhang et. al. include this issue in their performance diagnosis of the Hyperledger Fabric platform \cite{zhang20} while companies like Boxer Construction Analysts and Robinson Credit Company have implemented independent solutions to deal with this issue \cite{18_fabric17}}. Overall, Tendermint seemed to be the best in terms of throughput, and Clique in terms of average latency. Raft may perform better if Hyperledger Fabric is configured to deal with larger loads.

\begin{figure}[tbp]
  \centering
  \includegraphics[width=\hsize]{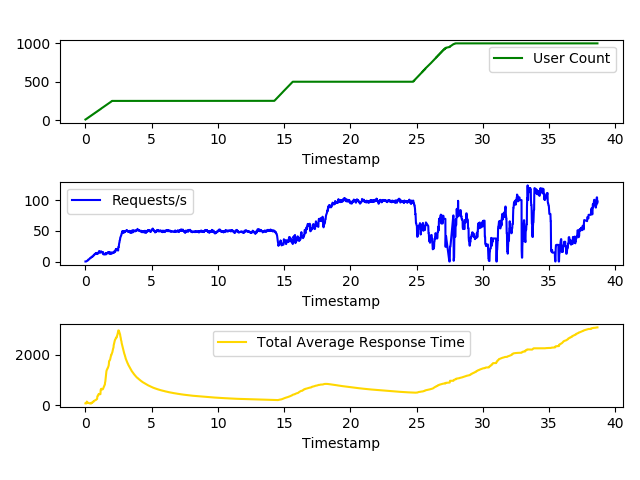}
  \caption{Blockchain application load test (PBFT)}
  \label{fig:app-lt-pbft}
\end{figure}

\begin{figure}[tbp]
  \centering
  \includegraphics[width=\hsize]{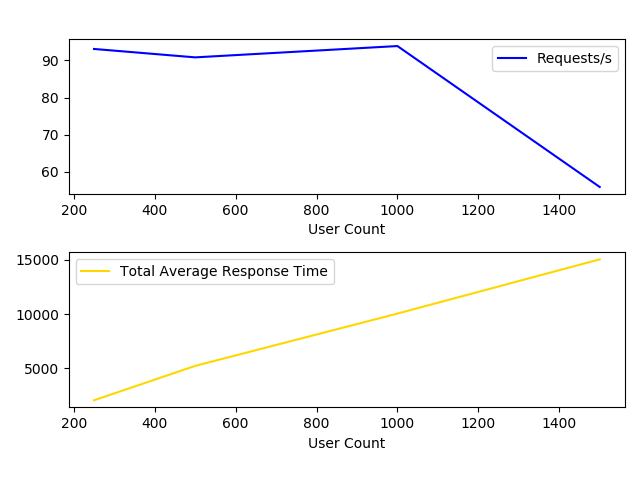}
  \caption{Blockchain application load test (Tendermint)}
  \label{fig:app-lt-tend}
\end{figure}

\begin{figure}[tbp]
  \centering
  \includegraphics[width=\hsize]{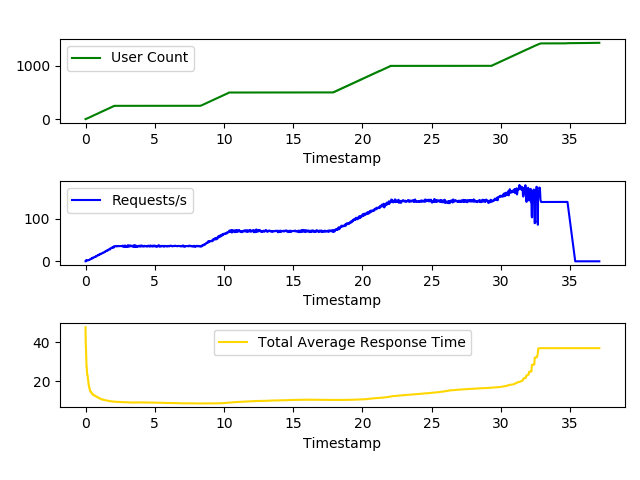}
  \caption{Blockchain application load test (Clique)}
  \label{fig:app-lt-cliq}
\end{figure}

\begin{figure}[tbp]
  \centering
  \includegraphics[width=\hsize]{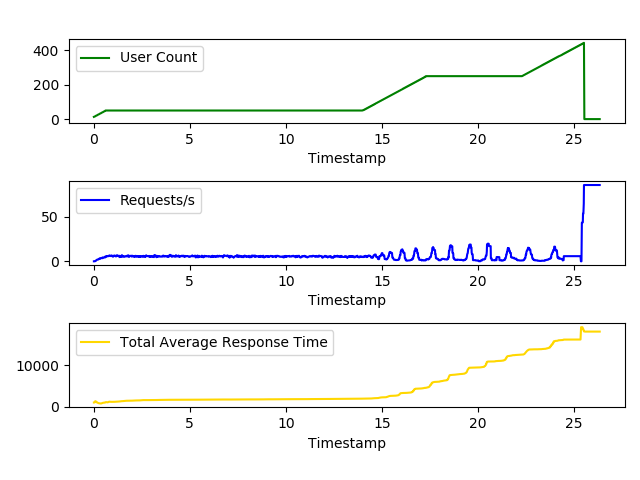}
  \caption{Blockchain application load test (Raft)}
  \label{fig:app-lt-raft}
\end{figure}

The load tests for each application were carried out until the application crashed, or performance degraded visibly. PBFT (Fig.~\ref{fig:app-lt-pbft}) did well till the load reached 1000 users, after which it quickly degraded. PBFT's throughput and average latency fluctuated when the load was changing but stabilized once the load stabilized. The Tendermint (Fig.~\ref{fig:app-lt-tend}) and Clique (Fig.~\ref{fig:app-lt-cliq}) applications showed the best performance under load - both crashed once they hit 1500 concurrent users causing system performance to degrade. Raft (Fig.~\ref{fig:app-lt-raft}) performed the worst under load. As discussed, Hyperledger Fabric's inability to naturally handle large loads explains why the performance is stable at lower loads but starts degrading/oscillating before even 250 users were spawned.

One final note on Tendermint explains why its average latency is so high compared to the other applications. Each account registered on the Tendermint network has an account number and a sequence number. The sequence number is incremented by the app every time the account makes a transaction. However, the internal copy of this sequence number only changes once the blockchain state is updated. While processing new transactions, the sequence number of the sending account is checked against its internal copy. If the two values do not match, the transaction is canceled. In other words, the application cannot accept new transactions from a given account, until the account's last transaction has been accepted (committed). Given a large number of concurrent users, each one ends up waiting for older transactions to be committed, which affects the average latency of the application.

\begin{table*}[tbp]
\caption{Blockchain applications: average performance metrics while chaos testing}\label{tab:ft}
\centering
\begin{adjustbox}{width=1\textwidth}
\begin{tabular}{|@{\vrule width0ptheight9pt\enspace}l|c|c|c|c|c|c|c|c|}\hline

\hfil\bf Protocol&\bf Metric&\bf baseline&\bf delay (100ms)& \bf loss (15\%)&\bf delay+loss&\bf corrupted (50\%)&\bf corrupted+delay+loss&\bf paused (50\%)\\\hline

\multirow{2}{*}{PBFT}&Throughput(tx/s)&50&17.5&16.2&24.78&10.5&16.5&4.9\\\cline{2-9}
&Median Latency(ms)&18&4463&20.88&4475&2055&4513&Null\\\hline
\multirow{2}{*}{Clique}&Throughput(tx/s)&27.3&28&28.5&28.5&25.76&24&5\\\cline{2-9}
&Median Latency (ms)&6&105&6&110&7&103&Null\\\hline
\multirow{2}{*}{Raft}&Throughput (tx/s)&5.8&5&4.8&3.75&3.82&3.55&2.33\\\cline{2-9}
&Median Latency (ms)&1766&3150&3300&5100&6271&6430&18500\\\hline

\end{tabular}
\end{adjustbox}
\end{table*}

\begin{figure}[tbp]
  \centering
  \includegraphics[width=\hsize]{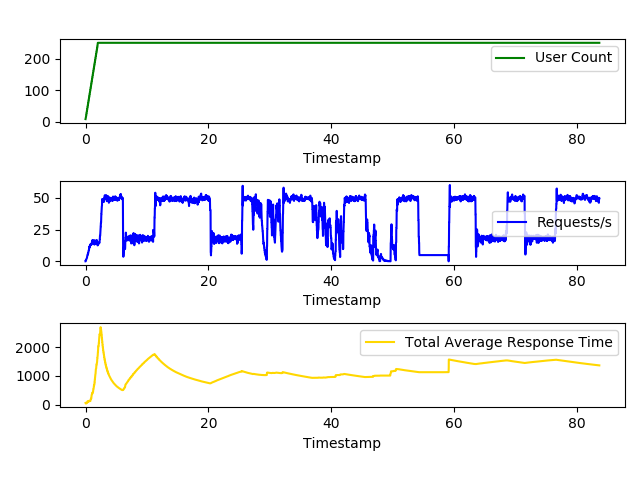}
  \caption{Blockchain application fault tolerance (PBFT)}
  \label{fig:app-ft-pbft}
\end{figure}

\begin{figure}[tbp]
  \centering
  \includegraphics[width=\hsize]{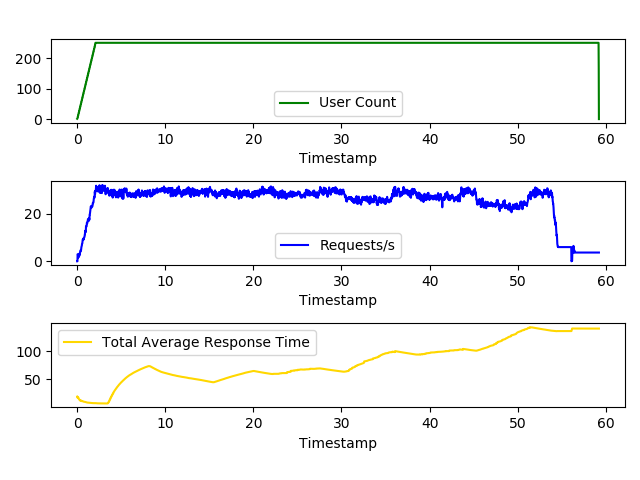}
  \caption{Blockchain application fault tolerance (Clique)}
  \label{fig:app-ft-cliq}
\end{figure}

\begin{figure}[tbp]
  \centering
  \includegraphics[width=\hsize]{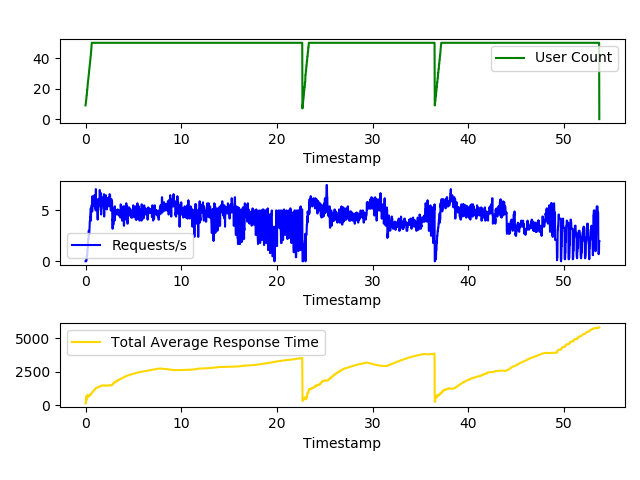}
  \caption{Blockchain application fault tolerance (Raft)}
  \label{fig:app-ft-raft}
\end{figure}

The chaos tests for each application were conducted at the same constant load as their respective baseline tests. The faults introduced during the test were (in order): delay, loss, delay and loss, corrupted messages from a single node, corrupted messages from half the network, corrupted messages (1 node) with delay and loss, corrupted messages (half network) with delay and loss, paused nodes. Corrupting outbound messages is similar to byzantine activity since each node receives contradicting messages. Similarly, pausing nodes is similar to simulating crash failures. The metric values when certain network faults were injected are specified in Table~\ref{tab:ft}. Figs.~\ref{fig:app-ft-pbft}-\ref{fig:app-ft-raft} depict the entire test during which the faults were simulated consecutively. In these test runs, following each fault mentioned above, the network was returned to normal conditions for an equal period before injecting the next fault. This can be observed in Fig.~\ref{fig:app-ft-pbft} where throughput returns to the baseline periodically. The throughput in these plots can be compared to the throughput in Table~\ref{tab:ft}. However, the latency in Table~\ref{tab:ft} refers to the median latency at each instant during the test, while the latency in Figs.~\ref{fig:app-lt-pbft}-\ref{fig:app-ft-raft} represents a running average of the latency throughout the entire test run.

Entries with a 'Null' value in Table~\ref{tab:ft} signify that no data was available for that test period. This was usually accompanied by short spikes where the latency metric degraded heavily. While the median response time (latency) may remain relatively low during the spike, the maximum response time shot up. For instance, when half the network was paused, the maximum response time degraded to 300000 ms in PBFT and 28000 ms in Clique. Apart from these short spikes, there was no data for latency during the periods in question. Pausing half the network nodes had the most drastic effect on performance compared to the other faults simulated. One noteworthy observation is that network faults affected the throughput of PBFT drastically but had very little effect on Clique's throughput. On the other hand, the average latency of Clique and PBFT did not change drastically, whereas Raft's average latency degraded continuously as different network faults were added and removed from the network. 

The Fabric application could not handle the test very well and crashed thrice, hence the drops in the plots of Fig.~\ref{fig:app-ft-raft}. This figure consists of three separate tests whose results were combined. The throughput plot for Raft looks like it fluctuates a lot, but this is due to the scale of the y-axis, and in reality, the extremes are not separated by much at the baseline. Similar to Clique, the faults affected Raft's latency more than its throughput. It can also be seen that Raft handled network delay or loss well compared to other faults.

\medskip

\subsection{Final Thoughts}
\textcolor{black}{Although the application results in Table~\ref{tab:ft} and the stochastic model results in Figs.~\ref{fig:mod-tp} and \ref{fig:mod-lat} are not comparable in terms of magnitude, they follow the same overall trends. For example, the throughput of PBFT is best with only delay, followed by corrupted messages with delay, and then corrupted messages without delay. Fig.~\ref{fig:mod-tp} gives the same relative order. The stochastic models show the baseline throughput performance follows the following order from best to worst: PBFT, Clique, Tendermint, Raft. The blockchain application results follow the following order: Tendermint, PBFT, Clique, Raft. This shows that the Tendermint application performed much better and the PBFT application much worse than the models predicted. Similarly, the models predicted the following order for average latency from best to worst: Clique, Raft, Tendermint, PBFT; while the application results followed the following order: Clique, PBFT, Raft, Tendermint. Here, PBFT did much better than the models predicted, while the others performed as expected.}

\textcolor{black}{It is important to also consider the role of platform architecture in these results. As discussed, Hyperledger Fabric (Raft) and Tendermint (Tendermint) follow specific rules that have an adverse effect on application performance. Since the models did not consider the account sequence numbers, they could not have predicted the degradation in Tendermint's latency. Similarly, if the Fabric application were built to handle a larger load as in \cite{18_fabric17}, it would improve the application's latency results. However, this needs to be verified. Overall, the models give a good understanding of how different protocols handle load and network faults. They also give a decent overview of the protocols' relative performance; however, it must be kept in mind that in addition to consensus protocols, blockchain platforms play an essential role in the performance of blockchain applications as well.}


\section{Conclusion}
\label{sec:conc}
\textcolor{black}{In this paper, we summarized our experiences in evaluating the performance of permissioned blockchain platforms using stochastic models and blockchain deployments. We applied chaos engineering principles to observe the performance of consensus algorithms and blockchain applications in faulty production environments. We observed how the performance characteristics changed for each protocol and how their performance changed as a result of user/transaction load and stochastic failures. It is clear that the choice of consensus algorithm affects system performance.} 

\textcolor{black}{The results show that the stochastic models and blockchain applications gave similar relative results amongst the consensus protocols and different faulty environments. Stochastic models can be used as a tool to filter out consensus protocols based on metrics like the performance metrics used in this work. We found that Tendermint and Clique were able to handle load better than PBFT and Raft, and Clique also maintained its throughput in faulty environments. However, Tendermint and PBFT showed a better throughput overall while Raft performed the worst. In addition, our experiments with blockchain platforms showed that the choice of blockchain platform plays an important role too. Therefore, no final decision must be made based solely on stochastic modeling experiments. It also indicates that if two different blockchain platforms use the same blockchain algorithm or protocol, we should not assume that they will have similar performance. This can be due to restrictive architecture as in Cosmos (Tendermint) or extra effort as in Hyperledger Fabric (Raft). Eventually, it is the choice of consensus protocol as well as the platform that decides the performance of a blockchain system.} 

\textcolor{black}{In the future, we plan to extend our chaos testing scenarios to design a complete chaos test suite for blockchain applications. We will investigate the reliability of more complex blockchain applications (functional scalability) in the presence of failures and investigate the impact of failure on geographic scalability. We will also work on standardizing the modeling process to make it easier to build models for a larger variety of consensus protocols using reusable blocks of code. Finally, we are interested in investigating the overhead introduced by different blockchain platforms, particularly the platforms that use the same consensus algorithms or protocols.}


\section{Acknowledgments}
\textcolor{black}{This project was supported in part by collaborative research funding from the National Research Council (NRC) of Canada’s Artificial Intelligence for Logistics (AI4L) Program.  We thank our colleagues from the NRC of Canada, who provided insight and expertise that greatly assisted the research. However, they may not agree with all of the interpretations/conclusions of this paper.}

\appendices
\section{Code Snippet}
\subsection{Stochastic Modelling}
The following code snippets demonstrate the central aspects of building stochastic models. All the code in this section is written using the pyCATSHOO modelling tool in Python. 

Code Snippet \ref{fig:code-mod-mc} shows how message boxes are created. A message box is a communication channel between two components which allows them to transfer values to and from eachother. In this snippet, the 'Peer' component has two message boxes with the 'Leader' component - one connection to receive values from the Leader and another to send values to the Leader. The Leader component will have the corresponding message box definitions to send values to the Peer and to receive values from it. 

\begin{lstlisting}[language=python, caption={Message channel definitions}, label={fig:code-mod-mc}]
# import from Leader 
messageBoxLead=self.addMessageBox("MB-fromProposer")
messageBoxLead.addImport(self.b_blockId, "blockId")
messageBoxLead.addImport(self.b_blockNum, "blockNum")
messageBoxLead.addImport(self.b_blockTs, "blockTimestamp")
messageBoxLead.addImport(self.b_totTx, "blockTotalTrans")
messageBoxLead.addImport(self.l_msg, "leaderMessage")
messageBoxLead.addImport(self.b_blockHash, "hashSequence")
messageBoxLead.addImport(self.l_propUID, "proposerUID")
# export to Leader
messageBoxLead2=self.addMessageBox("MB-toProposer")
messageBoxLead2.addExport(self.viewChanged, "viewChanged")
messageBoxLead2.addExport(self.isMal, "isMalicious")
messageBoxLead2.addExport(self.isBen, "isBenign")
messageBoxLead2.addExport(self.updateBlockTx, "updateBlockTx")
messageBoxLead2.addExport(self.invalid, "decidedInvalid")
messageBoxLead2.addExport(self.inWaiting, "inWaiting")
# import from Counter
messageBoxCounter1=self.addMessageBox("MB-fromCounter")
messageBoxCounter1.addImport(self.c_prevMaj, "prevMajorityReached")
messageBoxCounter1.addImport(self.c_precMaj, "precMajorityReached")
messageBoxCounter1.addImport(self.c_blockInvalid, "blocksInvalid")
# export to Counter
messageBoxCounter2=self.addMessageBox("MB-toCounter")
messageBoxCounter2.addExport(self.propRecd, "msgReceived")
messageBoxCounter2.addExport(self.prevUID, "prevoteUID")
messageBoxCounter2.addExport(self.precUID, "precommitUID")
messageBoxCounter2.addExport(self.blockInvalid, "invalidBlock")
\end{lstlisting}

In Code Snippet \ref{fig:code-mod-aut1} the functional automaton is created within the Peer component and the automaton states as well as the state transitions are defined. Each state transition consists of three steps: defining the transition name, checking the transition condition, and setting the target state. 

\begin{lstlisting}[language=python, caption={The peer state machine}, label={fig:code-mod-aut1}]
self.FuncAutomaton = self.addAutomaton("FunctionalAutomaton")
self.stateStart = self.addState("FunctionalAutomaton", "START", 0)
self.stateWaiting = self.addState("FunctionalAutomaton", "WAITING", 1)
self.stateProposed = self.addState("FunctionalAutomaton", "PROPOSE", 2)
self.statePrevoted = self.addState("FunctionalAutomaton", "PREVOTE", 3)
self.statePrecommitted = self.addState("FunctionalAutomaton", "PRECOMMIT", 4)
self.setInitState("START")
# State transitions
# Start         ->      Waiting
trans = self.stateStart.addTransition("start")
trans.setCondition("startTimer")
trans.addTarget(self.stateWaiting, Pyc.TTransType.trans)
# Waiting       ->      Start
trans = self.stateWaiting.addTransition("timedout")
trans.setCondition("timedOut")
trans.addTarget(self.stateStart, Pyc.TTransType.trans)
# Waiting       ->      Proposed
trans = self.stateWaiting.addTransition("proposed")
trans.setCondition("proposeRecd")
trans.addTarget(self.stateProposed, Pyc.TTransType.trans)
# Proposed      ->      Start
trans = self.stateProposed.addTransition("invalid")
trans.setCondition("invalidBlock")
trans.addTarget(self.stateStart, Pyc.TTransType.trans)
# Proposed      ->      Start
trans = self.stateProposed.addTransition("prevoteTimedout")
trans.setCondition("timedOut")
trans.addTarget(self.stateStart, Pyc.TTransType.trans)
# Proposed      ->      Prevoted
trans = self.stateProposed.addTransition("prevote")
trans.setCondition("majorityReached")
# trans.setDistLaw(Pyc.TLawType.expo, 3)
trans.addTarget(self.statePrevoted, Pyc.TTransType.trans)
# Prevoted      ->      Start
trans = self.statePrevoted.addTransition("precommitTimeout")
trans.setCondition("timedOut")
trans.addTarget(self.stateStart, Pyc.TTransType.trans)
# Prevoted      ->      Precommitted
trans = self.statePrevoted.addTransition("precommit")
trans.setCondition("majorityReached")
# trans.setDistLaw(Pyc.TLawType.expo, 3)
trans.addTarget(self.statePrecommitted, Pyc.TTransType.trans)
# Precommitted  ->      Start
trans = self.statePrecommitted.addTransition("finished")
trans.addTarget(self.stateStart, Pyc.TTransType.trans)
\end{lstlisting}

Code Snippet \ref{fig:code-mod-fn} defines what happens when a new block is created by any given peer in the blockchain model network. The important steps are making a note of the new length of the peer's blockchain and collecting performance-related information for the new block. Performance-related information includes the block's latency and the number of transactions in the block. 

\begin{lstlisting}[language=python, caption={addToChain() function}, label={fig:code-mod-fn}]
def addToChain(self):
	global PEER_LIMITER, PEER_CHAINLEN
	# add block to the chain 
	self.numberOfBlocks.setIValue(self.numberOfBlocks.value() + 1)
	self.viewChanged.setBValue(True)
	print(self.numberOfBlocks.value())
	PEER_CHAINLEN[self.uId.value()-1] = self.numberOfBlocks.value()
	if PEER_LIMITER == 0:
		global ADDED_BLOCK_TX
		ADDED_BLOCK_TX.append(TOT_BLOCKS)
		PEER_LIMITER += 1
	global NUM_BLOCKS, START_TIME_LIST, LATENCY_LIST
	NUM_BLOCKS = self.numberOfBlocks.value()
	if (len(LATENCY_LIST) < self.l.value() + self.b_totTx.iValue(0)):
		for t in range(self.b_totTx.iValue(0)):
			LATENCY_LIST.append(time.time() - START_TIME_LIST[t])
		self.l.setIValue(len(LATENCY_LIST))
		if len(START_TIME_LIST) > self.b_totTx.iValue(0):
			START_TIME_LIST = START_TIME_LIST[self.b_totTx.iValue(0):]
		else:
			START_TIME_LIST = []
\end{lstlisting}

In Code Snippet \ref{fig:code-mod-aut2} a second automaton is created for the Peer component. This automaton defines whether the Peer is malicious or benign at any given time. The state transitions for this automaton are different from the Functional Automaton. A benign peer turns malicious on the condition that the byzantine rate (percentage of malicious peers) defined at runtime is not violated. If this condition is fulfilled the peer will turn malicious probabilistically (based on a probability distribution). Similarly, a malicious peer can turn benign probabilistically. 

\begin{lstlisting}[language=python, caption={Peer automaton for byzantine behaviour}, label={fig:code-mod-aut2}]
self.TypeAutomaton = self.addAutomaton("TypeAutomaton")
self.stateMalicious = self.addState("TypeAutomaton", "MALICIOUS", 0)
self.stateBenign = self.addState("TypeAutomaton", "BENIGN", 1)
self.setInitState("BENIGN")
# State transitions
# Benign        ->      Malicious
trans = self.stateBenign.addTransition("faulty")
trans.setCondition("checkMalicious")
trans.setDistLaw(Pyc.TLawType.defer, 5)
trans.setInterruptible(True)
trans.addTarget(self.stateMalicious, Pyc.TTransType.trans)
# Malicious     ->      Benign
trans = self.stateMalicious.addTransition("benign")
trans.setDistLaw(Pyc.TLawType.defer, 50)
trans.addTarget(self.stateBenign, Pyc.TTransType.trans)
\end{lstlisting}

\subsection{Blockchain Testbeds}
The following code snippets summarise how the various blockchain testbed applications are created. Code Snippet \ref{fig:code-saw} is the YAML file used to create the docker image for each Sawtooth validator. It defines the communication ports as well as the initialization bash commands.  

\begin{lstlisting}[language=python, caption={Hyperledger Sawtooth validator configuration}, label={fig:code-saw}]
validator-1:
image: hyperledger/sawtooth-validator:chime
container_name: sawtooth-validator-default-1
expose:
  - 4004
  - 5050
  - 8800
volumes:
  - pbft-shared:/pbft-shared
command: |
  bash -c "
    if [ -e /pbft-shared/validators/validator-1.priv ]; then
      cp /pbft-shared/validators/validator-1.pub /etc/sawtooth/keys/validator.pub
      cp /pbft-shared/validators/validator-1.priv /etc/sawtooth/keys/validator.priv
    fi &&
    if [ ! -e /etc/sawtooth/keys/validator.priv ]; then
      sawadm keygen
      mkdir -p /pbft-shared/validators || true
      cp /etc/sawtooth/keys/validator.pub /pbft-shared/validators/validator-1.pub
      cp /etc/sawtooth/keys/validator.priv /pbft-shared/validators/validator-1.priv
    fi &&
    sawtooth keygen my_key &&
    sawtooth-validator -vv \
      --endpoint tcp://validator-1:8800 \
      --bind component:tcp://eth0:4004 \
      --bind consensus:tcp://eth0:5050 \
      --bind network:tcp://eth0:8800 \
      --scheduler parallel \
      --peering static \
      --maximum-peer-connectivity 10000 \
      --peers tcp://validator-0:8800
  "
\end{lstlisting}

Code Snippet \ref{fig:code-cos-handle} shows the definition of the Cosmos handler which determines what happens when a 'Create Transaction' command is sent to a peer. Basic validations are done and the transaction information is routed to the appropriate Keeper function which in turn updates the blockchain state. 

\begin{lstlisting}[language=python, caption={Cosmos handler}, label={fig:code-cos-handle}]
func handleMsgCreateTrans(ctx sdk.Context, k keeper.Keeper, msg types.MsgCreateTrans) (*sdk.Result, error) {
	var trans = types.Trans{
		Sender:     msg.Sender,
		Receiver:   msg.Receiver,
		Amount:     msg.Amount,
	}

	amt := strconv.Itoa(trans.Amount) + "token"
	payment, _ := sdk.ParseCoins(amt)
	if err := k.CoinKeeper.SendCoins(ctx, trans.Sender, trans.Receiver, payment); err != nil {
		return nil, err
	}
	k.CreateTrans(ctx, msg)

	return &sdk.Result{Events: ctx.EventManager().Events()}, nil
}
\end{lstlisting}

The Cosmos Keeper in Code Snippet \ref{fig:code-cos-keep} creates and adds the transaction to the blockchain. 

\begin{lstlisting}[language=python, caption={Cosmos keeper}, label={fig:code-cos-keep}]
func (k Keeper) CreateTrans(ctx sdk.Context, msg types.MsgCreateTrans) {
	// Create the trans
	count := k.GetTransCount(ctx)
	var trans = types.Trans{
		Sender:  msg.Sender,
		ID:       strconv.FormatInt(count, 10),
		Receiver: msg.Receiver,
		Amount:   msg.Amount,
	}

	store := ctx.KVStore(k.storeKey)
	key := []byte(types.TransPrefix + trans.ID)
	value := k.cdc.MustMarshalBinaryLengthPrefixed(trans)
	store.Set(key, value)

	// Update trans count
	k.SetTransCount(ctx, count+1)
}
\end{lstlisting}

The following snippets outline the settings in a Hyperledger Fabric network. Code Snippet \ref{fig:code-fab-ca} creates a docker image for the Hyperledger Fabric certificate authority. The ports and the start-up bash commands are defined. 

\begin{lstlisting}[language=python, caption={Docker configuration for Hyperledger Fabric's certificate authority}, label={fig:code-fab-ca}]
ca_org1:
    image: hyperledger/fabric-ca
    environment:
      - FABRIC_CA_HOME=/etc/hyperledger/fabric-ca-server
      - FABRIC_CA_SERVER_CA_NAME=ca.org1.example.com
      - FABRIC_CA_SERVER_TLS_ENABLED=true
      - FABRIC_CA_SERVER_PORT=7054
    ports:
      - "7054:7054"
    command: sh -c 'fabric-ca-server start -b admin:adminpw -d'
    volumes:
      - ./fabric-ca/org1:/etc/hyperledger/fabric-ca-server
    container_name: ca.org1.example.com
    hostname: ca.org1.example.com
    networks:
      - test
\end{lstlisting}


Code Snippet \ref{fig:code-fab-ord} creates a docker image for the network orderers (i.e. the peers). In addition to the port definitions, each orderer's environment is populated with certificates, keys and other necessary information. 

\begin{lstlisting}[language=python, caption={Docker configuration for Hyperledger Fabric's orderers}, label={fig:code-fab-ord}]
orderer.example.com:
    container_name: orderer.example.com
    image: hyperledger/fabric-orderer:2.1
    dns_search: .
    environment:
      - ORDERER_GENERAL_LOGLEVEL=info
      - FABRIC_LOGGING_SPEC=INFO
      - ORDERER_GENERAL_LISTENADDRESS=0.0.0.0
      - ORDERER_GENERAL_GENESISMETHOD=file
      - ORDERER_GENERAL_GENESISFILE=/var/hyperledger/orderer/genesis.block
      - ORDERER_GENERAL_LOCALMSPID=OrdererMSP
      - ORDERER_GENERAL_LOCALMSPDIR=/var/hyperledger/orderer/msp
      - ORDERER_GENERAL_TLS_ENABLED=true
      - ORDERER_GENERAL_TLS_PRIVATEKEY=/var/hyperledger/orderer/tls/server.key
      - ORDERER_GENERAL_TLS_CERTIFICATE=/var/hyperledger/orderer/tls/server.crt
      - ORDERER_GENERAL_TLS_ROOTCAS=[/var/hyperledger/orderer/tls/ca.crt]
      - ORDERER_KAFKA_VERBOSE=true
      - ORDERER_GENERAL_CLUSTER_CLIENTCERTIFICATE=/var/hyperledger/orderer/tls/server.crt
      - ORDERER_GENERAL_CLUSTER_CLIENTPRIVATEKEY=/var/hyperledger/orderer/tls/server.key
      - ORDERER_GENERAL_CLUSTER_ROOTCAS=[/var/hyperledger/orderer/tls/ca.crt]
      - ORDERER_METRICS_PROVIDER=prometheus
      - ORDERER_OPERATIONS_LISTENADDRESS=0.0.0.0:8443
      - ORDERER_GENERAL_LISTENPORT=7050
    working_dir: /opt/gopath/src/github.com/hyperledger/fabric/orderers
    command: orderer
    ports:
      - 7050:7050
      - 8443:8443
    networks:
      - test
    volumes:
      - ./channel/genesis.block:/var/hyperledger/orderer/genesis.block
      - ./channel/crypto-config/ordererOrganizations/example.com/orderers/orderer.example.com/msp:/var/hyperledger/orderer/msp
      - ./channel/crypto-config/ordererOrganizations/example.com/orderers/orderer.example.com/tls:/var/hyperledger/orderer/tls
\end{lstlisting}


A Hyperledger Fabric channel's creation is described in Code Snippet \ref{fig:code-fab-chan}. The channel artifacts are generated at the specified path. Once the channel is created, each peer is then able to join the channel.

\begin{lstlisting}[language=python, caption={Hyperledger Fabric channel preparation}, label={fig:code-fab-chan}]
createChannel(){
    rm -rf ./channel-artifacts/*
    setGlobalsForPeer0Org1
    
    peer channel create -o localhost:7050 -c $CHANNEL_NAME \
    --ordererTLSHostnameOverride orderer.example.com \
    -f ./artifacts/channel/${CHANNEL_NAME}.tx --outputBlock ./channel-artifacts/${CHANNEL_NAME}.block \
    --tls $CORE_PEER_TLS_ENABLED --cafile $ORDERER_CA
}

joinChannel(){
    setGlobalsForPeer0Org1
    peer channel join -b ./channel-artifacts/$CHANNEL_NAME.block
    
    setGlobalsForPeer1Org1
    peer channel join -b ./channel-artifacts/$CHANNEL_NAME.block

    setGlobalsForPeer0Org2
    peer channel join -b ./channel-artifacts/$CHANNEL_NAME.block

    setGlobalsForPeer1Org2
    peer channel join -b ./channel-artifacts/$CHANNEL_NAME.block
    
    setGlobalsForPeer0Org3
    peer channel join -b ./channel-artifacts/$CHANNEL_NAME.block

    setGlobalsForPeer1Org3
    peer channel join -b ./channel-artifacts/$CHANNEL_NAME.block
}
\end{lstlisting}


In Code Snippet \ref{fig:code-fab-cc} the chaincode script is archived and packaged in a way that it is usable by each peer. The final step in chaincode preparation is to install the chaincode into each peer's environment. 

\begin{lstlisting}[language=python, caption={Hyperledger Fabric chaincode preparation}, label={fig:code-fab-cc}]
packageChaincode() {
    rm -rf ${CC_NAME}.tar.gz
    setGlobalsForPeer0Org1
    peer lifecycle chaincode package ${CC_NAME}.tar.gz \
        --path ${CC_SRC_PATH} --lang ${CC_RUNTIME_LANGUAGE} \
        --label ${CC_NAME}_${VERSION}
    echo "===================== Chaincode is packaged ===================== "
}
# packageChaincode

installChaincode() {
    setGlobalsForPeer0Org1
    peer lifecycle chaincode install ${CC_NAME}.tar.gz
    echo "===================== Chaincode is installed on peer0.org1 ===================== "

    setGlobalsForPeer1Org1
    peer lifecycle chaincode install ${CC_NAME}.tar.gz
    echo "===================== Chaincode is installed on peer1.org1 ===================== "

    setGlobalsForPeer0Org2
    peer lifecycle chaincode install ${CC_NAME}.tar.gz
    echo "===================== Chaincode is installed on peer0.org2 ===================== "

    setGlobalsForPeer1Org2
    peer lifecycle chaincode install ${CC_NAME}.tar.gz
    echo "===================== Chaincode is installed on peer1.org2 ===================== "

    setGlobalsForPeer0Org3
    peer lifecycle chaincode install ${CC_NAME}.tar.gz
    echo "===================== Chaincode is installed on peer0.org3 ===================== "

    setGlobalsForPeer1Org3
    peer lifecycle chaincode install ${CC_NAME}.tar.gz
    echo "===================== Chaincode is installed on peer1.org3 ===================== "
}
\end{lstlisting}


Code Snippet \ref{fig:code-fab-cccfn} shows the main chaincode function which performs necessary validations and handles the settlement of accounts on the blockchain. Once the transaction is validated, the state of the blockchain can be updated. 

\begin{lstlisting}[language=python, caption={Hyperledger Fabric chaincode function}, label={fig:code-fab-cccfn}]
func (s *SmartContract) UpdateAccountBalance(ctx contractapi.TransactionContextInterface, fromAccID string, toAccID string, amt uint64) (string, error) {

	if len(fromAccID) == 0 {
		return "", fmt.Errorf("Please pass the correct account id to send from")
	}

	if len(toAccID) == 0 {
		return "", fmt.Errorf("Please pass the correct account id to send to")
	}

	// sending account
	accAsBytesFrom, err := ctx.GetStub().GetState(fromAccID)
	if err != nil {
		return "", fmt.Errorf("Failed to get sender account data. %s", err.Error())
	}
	if accAsBytesFrom == nil {
		return "", fmt.Errorf("%s does not exist", fromAccID)
	}

	fromAccount := new(Account)
	_ = json.Unmarshal(accAsBytesFrom, fromAccount)

	// receiving account
	accAsBytesTo, err := ctx.GetStub().GetState(toAccID)
	if err != nil {
		return "", fmt.Errorf("Failed to get receiver account data. %s", err.Error())
	}
	if accAsBytesTo == nil {
		return "", fmt.Errorf("%s does not exist", toAccID)
	}

	toAccount := new(Account)
	_ = json.Unmarshal(accAsBytesTo, toAccount)

	// update account balances
	fromAccount.Balance = fromAccount.Balance - amt
	toAccount.Balance = toAccount.Balance + amt

	// to account
	accAsBytesTo, err = json.Marshal(toAccount)
	if err != nil {
		return "", fmt.Errorf("Failed while marshling receiver's account. %s", err.Error())
	}
	err = ctx.GetStub().PutState(toAccount.ID, accAsBytesTo)
	if err != nil {
		return "", fmt.Errorf("Failed to set update receiver's account. %s", err.Error())
	}

	// from account
	accAsBytesFrom, err = json.Marshal(fromAccount)
	if err != nil {
		return "", fmt.Errorf("Failed while marshling sender's account. %s", err.Error())
	}

	//  txId := ctx.GetStub().GetTxID()

	return ctx.GetStub().GetTxID(), ctx.GetStub().PutState(fromAccount.ID, accAsBytesFrom)
}
\end{lstlisting}


Code Snippet \ref{fig:code-fab-app} shows the function invoked when a new transaction is initiated on the blockchain. This Node.js application acts as an interface between the frontend/user and the blockchain backend. 

\begin{lstlisting}[language=python, caption={Hyperledger Fabric app endpoint to invoke a transaction}, label={fig:code-fab-app}]
app.post('/channels/:channelName/chaincodes/:chaincodeName', async function (req, res) {
    try {
        logger.debug('==================== INVOKE ON CHAINCODE ==================');
        var peers = req.body.peers;
        var chaincodeName = req.params.chaincodeName;
        var channelName = req.params.channelName;
        var fcn = req.body.fcn;
        var args = req.body.args;
        var transient = req.body.transient;
        console.log(`Transient data is ;${transient}`)
        logger.debug('channelName  : ' + channelName);
        logger.debug('chaincodeName : ' + chaincodeName);
        logger.debug('fcn  : ' + fcn);
        logger.debug('args  : ' + args);
        if (!chaincodeName) {
            res.json(getErrorMessage('\'chaincodeName\''));
            return;
        }
        if (!channelName) {
            res.json(getErrorMessage('\'channelName\''));
            return;
        }
        if (!fcn) {
            res.json(getErrorMessage('\'fcn\''));
            return;
        }
        if (!args) {
            res.json(getErrorMessage('\'args\''));
            return;
        }
        let message = await invoke.invokeTransaction(channelName, chaincodeName, fcn, args, req.username, req.orgname, transient);
        console.log(`message result is : ${message}`)
        const response_payload = {
            result: message,
            error: null,
            errorData: null
        }
        res.send(response_payload);
    } catch (error) {
        const response_payload = {
            result: null,
            error: error.name,
            errorData: error.message
        }
        res.send(response_payload)
    }
});
\end{lstlisting}

\subsection{Performance Measurement}
This section contains code snippets which demonstrate how performance was measured in the blockchain testbed applications. Code Snippet \ref{fig:code-fab-perf} shows how a transaction is created and posted at the appropriate endpoint of the Hyperledger Fabric application. The tool used (Locust) compiles the performance metrics and presents them in a GUI which has been recreated in Section \ref{sec:results}.

\begin{lstlisting}[language=python, caption={Locust file to measure the performance of the Hyperledger Fabric app}, label={fig:code-fab-perf}]
txn_header = {
    'Accept': '*/*',
    'Connection': 'keep-alive',
    'Authorization': 'Bearer {}'.format(REGISTERED_USERS[auth_user][1]),
    'Content-Type': 'application/json'
}
# set urls 
posturl = '/channels/' + self.channel_name + '/chaincodes/' + self.chaincode_name
sendurl = self.baseurl + posturl

# post asset transfer 
with self.client.post(sendurl, data=json.dumps(payload), headers=txn_header,
                    catch_response=True) as response:
    # if response.status_code < 400:
    try:
        if not response.json()['error']:
            print("Transfer success")
            response.success()
        else:
            response.failure("Transfer failed with error: %s" % response.json()['error'])
    except JSONDecodeError:
        response.failure("Request failed with error: No valid responses from any peers.")
\end{lstlisting}


Code Snippet \ref{fig:code-cos-perf1} shows how the Cosmos application is run multiple times in separate terminal sessions and the performance metrics calculated in each session are compiled together to give the average results. 

\begin{lstlisting}[language=python, caption={Measuring the performance of the Cosmos app - 1}, label={fig:code-cos-perf1}]
start_time = time.time()
# open terminals and run testfile
for i in range(num_terminals):
	if i == 0:
		os.system("tmux splitw -d -h 'python3 {} {}.txt'".format(testfile, 
				savefile+str(i)))
	elif i == (num_terminals-1):
		os.system("tmux splitw -d -h 'python3 {} {}.txt'".format(testfile, 
				savefile+str(i)))
		os.system("tmux select-layout even-horizontal")
	else:
		os.system("tmux new -s mysesh{} -d 'python3 {} {}.txt'".format(i, 
				testfile, savefile+str(i)))

# wait for savefiles to be created 
while len(file_count) != num_terminals:
	for i in range(num_terminals):
		if os.path.exists(savefile+str(i)+".txt") and (i not in file_count):
			file_count.append(i)
	# time.sleep(sleeptime)
tot_time = time.time() - start_time

# compile results
for i in range(num_terminals):
	with open(savefile+str(i)+".txt", 'r') as f:
		data = f.readlines()
		tot_tx.append(int(data[0].strip()))
		l.append(float(data[1].strip()))
		sr.append(float(data[2].strip()))
		tp.append(float(data[3].strip()))
	# delete extra files
	os.remove(savefile+str(i)+'.txt')

# prepare results to write to file
tp_str = ['{:.3f}'.format(x) for x in tp]
l_str = ['{:.3f}'.format(x) for x in l]
sr_str = ['{:.2f}'.format(x) for x in sr]
\end{lstlisting}


Code Snippet \ref{fig:code-cos-perf2} shows how the performance metrics are calculated within each terminal session. The code in Snippets \ref{fig:code-cos-perf1} and \ref{fig:code-cos-perf2} exist in separate files, Code Snippet \ref{fig:code-cos-perf2} sends its results to Snippet \ref{fig:code-cos-perf1} to be averaged out. 

\begin{lstlisting}[language=python, caption={Measuring the performance of the Cosmos app - 2}, label={fig:code-cos-perf2}]
while num_rounds < NUM_ROUNDS:
	# variables 
	tx_hash = 0
	flags = [False] * TRANS_PER_ROUND
	print("\nRound number {}/{}".format(num_rounds+1, NUM_ROUNDS))

	for i in range(TRANS_PER_ROUND):
		# send transaction 
		print("Round{}: {}/{}.".format(num_rounds+1, i+1, TRANS_PER_ROUND), end=" ")
		l_start_time = time.time()
		while tx_hash == 0:
			tx_hash = send_tx(recipients[random.choice(users)], 
					str(random.randint(5,50)),
					random.choice(users), wait_period)
		# confirm transaction is comitted
		c_start_time = time.time()
		while True:
			if time.time() - c_start_time > c_wait_time:
				print("Timedout.")
				break
			try:
				subprocess.run(['simpleAppcli', 'query', 'tx', tx_hash],
						stdout=subprocess.DEVNULL,
						stderr=subprocess.DEVNULL,
						check=True)
				print("Confirmed.")
				trans_success += 1
				latency_list.append(time.time() - l_start_time)
				break
			except:
				continue
	num_rounds += 1	
tot_time = time.time() - t_start_time
tp, l, sr = measure_performance(tot_time, latency_list, trans_success) 
\end{lstlisting}

\end{document}